\documentclass[pre,showpacs,preprint,superscriptaddress]{revtex4-1}
\usepackage{amssymb,graphicx,amsbsy,amsmath}

\begin{document}
\title{Theory of fads: Traveling-wave solution of evolutionary dynamics
in a one-dimensional trait space}
\author{Mi Jin Lee}
\affiliation{Department of Physics,
Sungkyunkwan University, Suwon 440-746, Korea}
\author{Su Do Yi}
\affiliation{Department of Physics, Pukyong National University, Busan 608-737,
Korea}
\author{Beom Jun Kim}
\email{beomjun@skku.edu}
\affiliation{Department of Physics,
Sungkyunkwan University, Suwon 440-746, Korea}
\author{Seung Ki Baek}
\email{seungki@pknu.ac.kr}
\affiliation{Department of Physics, Pukyong National University, Busan 608-737,
Korea}

\begin{abstract}
We consider an infinite-sized population where an infinite number of traits
compete simultaneously. The replicator equation with a diffusive term
describes time evolution of the probability distribution over the traits
due to selection and mutation on a mean-field level. We argue that this
dynamics can be expressed as a variant of the Fisher equation with
high-order correction terms. The equation has a traveling-wave solution,
and the phase-space method shows how the wave shape depends on the
correction. We compare this solution with empirical time-series data of
given names in Quebec, treating it as a descriptive model
for the observed patterns. Our model explains the reason that many names
exhibit a similar pattern of the rise and fall as time goes by.
At the same time, we have found that their dissimilarities are also
statistically significant.
\end{abstract}

\pacs{89.65.Cd,89.75.Kd,87.23.Ge}

\maketitle

\section{Introduction}
Sir Isaac Newton reportedly said that he could calculate the motion of
heavenly bodies but not the madness of people, when he had lost a fortune
in the South Sea Bubble.
Since his time, there have been vigorous attempts to apply the so-called
Newtonian approach to our society in order to understand the `madness of
people'.
It is interesting that some of modern economists have finally explained
bubbles as based on rational expectation~\cite{camerer,schaller},
although it does
not mean that bubbles are really under our control. At the same time, they
distinguish a fad from a bubble as the former originates solely from social
forces, which are harder to rationalize from an economic point of view.
Even if we take a more phenomenological viewpoint,
how this social influence organizes itself seems to remain largely
unpredictable.
A society adopts some practices and abandons some others constantly, but it
is neither a matter of practical use nor that of aesthetic superiority.
One might say that it is because something is `cool', but it hardly
explains anything but the unpredictability. It is another, perhaps a worse
kind of `madness' from a physicist's point of view.

\begin{figure}
\includegraphics[width=0.45\textwidth]{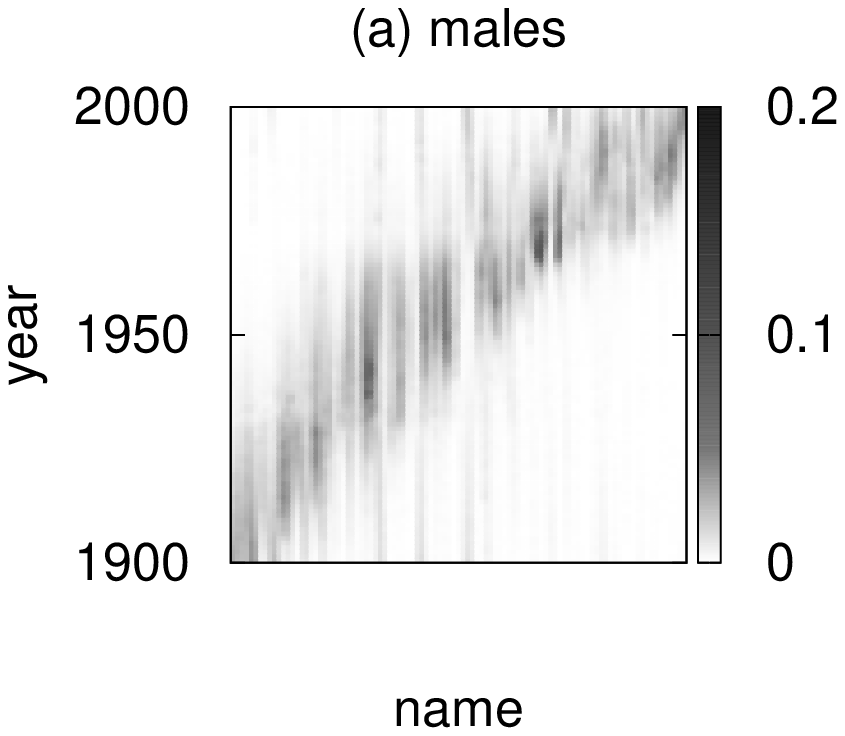}
\includegraphics[width=0.45\textwidth]{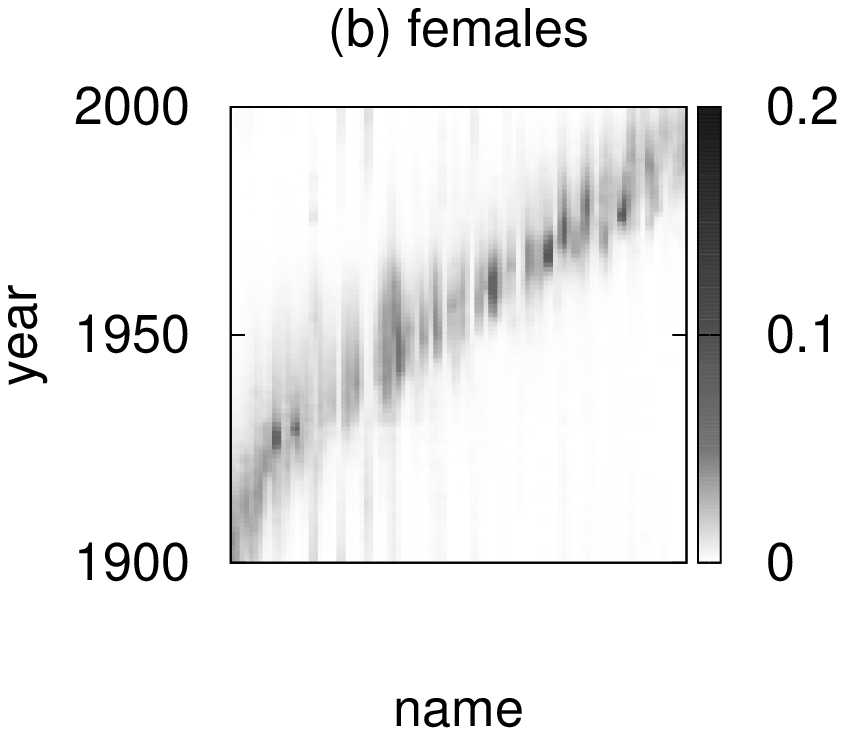}
\caption{ (a) Fractions of a hundred male given names and (b) those of
the same number of female given names in Quebec, from year $1900$ to $2000$.
The names are sorted in an ascending order of the mean time of prevalence
[Eq.~(\ref{eq:meant})].
}
\label{fig:wave}
\end{figure}

Given names are subject to fads~\cite{given,kessler}.
It is sharply contrasted to
the case of family names, whose dynamics is well defined by mathematical
models~\cite{zanette,reed,sk,rossi}.
If a certain given name prevails, on the other hand, it is simply because
it sounds cool, i.e., for no particular reason.
Note that being a fad in this context does not necessarily mean that
it is short-lived: For example,
Michael was the most popular name in the United States for about half
a century from the 1950s~\cite{comm}.
Our point is that such popularity is not explained by any of
its intrinsic properties, and we will call it a fad regardless of
the time scale.
Even if it is difficult to predict a fad, e.g., what will be the most
popular name for babies next year, one may still
expect a higher degree of regularity in the rise and fall of each given name.
Fortunately, there is an available data set, compiled by Duchesne, which
has recorded frequencies of $100$ major given names in the Canadian
province of Quebec for more than a
century~\cite{duchesne,statq}. It is basically a collection of names given to
a certain number of people each year, from which the fractions of the given
names are calculated. This collection is also called a corpus and
its size, i.e., the number of people surveyed, has been greater
than $2\times 10^3$ every year throughout the 20th century.
In Fig.~\ref{fig:wave}, the shade represents the fraction,
and the names are sorted according to the mean time of prevalence, defined as
\begin{equation}
\left<t \right> = \frac{\sum_t t f(t)}{\sum_t f(t)},
\label{eq:meant}
\end{equation}
where $t$ runs from year 1900 to 2000, and $f(t)$ is the fraction of
a particular name in year $t$.
This plot suggests the existence of a certain regular pattern.
That is, fads propagate with forming a traveling wave across the space of
names, from one name to another, and it is more clearly
seen for female names.
In addition, the narrower width of the name
distribution along the vertical axis in Fig.~\ref{fig:wave}(b)
indicates that the change is more rapid for girls' names.
This observation implies that the female-name dynamics is probably simpler,
in the sense that one only has to be concerned about how his or her
daughter's name will sound to others.
Based on this observation, we will construct a
descriptive model for the pattern in the female names.
While the existing models in Refs.~\onlinecite{given,kessler}
consider the dynamics of a single name interacting with the environment,
we are more interested in describing the competition
among many names that takes place simultaneously.
This many-body problem reduces to a reaction-diffusion system
written in terms of the frequency distribution of newborn children's
given names, yielding a traveling-wave solution.
By comparing the model with the empirical data, we will check whether
our picture captures essential parts of the dynamics.

This work is organized as follows. The next section presents our model,
formulated as a generalized version of the Fisher equation~\cite{fisher}.
Section~\ref{sec:data} then explains how we analyze the empirical data
and compare them with our model. In this section, the goodness of fit will
be quantitatively estimated with a statistical test. After discussing
the result in Sec.~\ref{sec:discussion}, we conclude this work.

\section{Model}

Our starting point is
adoption-exploration dynamics of idea spreading~\cite{inno}.
According to this theory,
we may imagine that each given name behaves as a
biological species, competing to win as many adopters as possible.
In other words, given names are assumed to obey certain evolutionary dynamics.
To formulate this dynamics in mathematical terms,
we need a number (or a set of numbers) to quantify
the competitiveness of each name relative to that of others.
Just as a species is said to have its own fitness~\cite{fit1,fit2}, therefore,
let us furthermore
assume that it is possible to assign a certain value of `fitness',
represented by a real number $x$, to each given name.
Our viewpoint is that
the names are arranged and indexed by this fitness measure, and a name with
$x$ is picked up by a newborn baby's parents from a certain probability
distribution function.
Note that our formulation mainly concerns how likely a name will be chosen for
a newborn baby, rather than the fraction of all the people carrying that name.
It is also important to remember that each given name has a fixed value of
$x$, which is independent of time $t$, so that we can
effectively map each $x$ to a different name.
For this reason,
one can think of $x$ as a position in the space of names, analogous to the
horizontal axis in Fig.~\ref{fig:wave},
although we do not actually have to determine
the values for the data set in our analysis as will be seen below.
So it is the \emph{fraction} of $x$, instead of its value, that changes with
time. Due to the growth of the fitter,
the distribution will move to higher values of $x$ in general.
As a result, a time series for a specific given name can be unimodal at best.
This might not be strictly true in Western countries~\cite{kessler}.
But in Quebec, only a
few percent of names succeed to come back. In our data set,
therefore, such recurrence as considered in Ref.~\onlinecite{kessler} can
rather be treated as exceptional.

Let us consider the following adoption-exploration dynamics~\cite{inno}:
\begin{equation}
\frac{\partial P}{\partial t} = k ( U - \left< U \right> ) P + D
\frac{\partial^2 P}{\partial x^2},
\label{eq:rd}
\end{equation}
where $P=P(x,t)$ is the probability density function at time $t$ for finding
a `species' of fitness $x \in (-\infty, +\infty)$, $U=U(x,t)$ is the payoff
that the species gains, and $\left< U \right>$ is the population
average of $U(x,t)$ defined as $\left<U \right> \equiv \int dx U(x,t) P(x,t)$.
The first term on the right-hand side (RHS) means that $P(x,t)$
has a relative growth rate proportional to $U(x,t)$, in a similar spirit
to the replicator equation in evolutionary biology~\cite{taylor,hof,rd}.
One may also refer to Ref.~\onlinecite{oech} for mathematical aspects of
a replicator equation on a continuous trait space.
The second term on the RHS describes exploratory activities for new traits
as diffusion along the $x$ axis.
The positive parameters $k$ and $D$ are thus called the rate of adoption and a
measure of exploration, respectively.
It is instructive to compare this with Eigen's phenomenological theory
of selection~\cite{eigen,ebeling}.
From our perspective, it is important that
this theory of selection can generally be applied to any information carriers
that reproduce themselves, including memes.
The theory starts with the following rate equation
for $P_i$, the concentration of an information carrier $i$ ($i=1,\ldots,n$):
\begin{equation}
\frac{d}{dt} P_i = F_i P_i - R_i P_i + \sum_{j} (\phi_{ij} P_j - \phi_{ji} P_i),
\end{equation}
where the first and second terms refer to self-instructed reproduction and removal
of the carrier, respectively, whereas the last summation contains all the other
production such as mutation.
As one of possible conditions for selection,
we may choose $\sum_i P_i = \mbox{const}$.
The reproduction and removal
can be specified further as follows:
Let us set $F_i = k_0 A_i Q_i$, where $k_0$ is a rate constant, $A_i$ is an
amplification factor, and $Q_i$ is a quality factor of precise reproduction.
Likewise, we set $R_i = k_0 B_i + \phi_{0}$, where $B_i$ is a decomposition
factor and $\phi_{0}$ is a dilution factor, which is controlled by the boundary
condition and assumed to be independent of $i$. We now obtain a more detailed
form of the rate equation:
\begin{equation}
\frac{d}{dt} P_i = k_0 \left( A_i Q_i - B_i \right) P_i
+ \sum_{j} (\phi_{ij} P_j - \phi_{ji} P_i) - \phi_{0} P_i.
\label{eq:eigen}
\end{equation}
If we sum up both sides of Eq.~(\ref{eq:eigen}) over $i$,
we find that
\begin{equation}
0 = \sum_{i} k_0 \left( A_i Q_i - B_i \right) P_i - \phi_{0} \sum_{i} P_i,
\label{eq:sumup}
\end{equation}
because $\sum_{i} \sum_{j} (\phi_{ij} P_j - \phi_{ji} P_i)$ identically
vanishes.
If we define a selective value as $U_i = A_i Q_i - B_i$,
Eq.~(\ref{eq:sumup}) means that
$\phi_0 = k_0 \sum_i U_i P_i / \sum_i P_i = k_0 \left< U \right>$.
Equation~(\ref{eq:eigen}) is thus rewritten as
\begin{equation}
\frac{d}{dt} P_i = k_0 \left( U_i - \left< U \right> \right) P_i
+ \sum_{j} (\phi_{ij} P_j - \phi_{ji} P_i).
\label{eq:disc}
\end{equation}
Supposing that the mutation rates $\phi_{ij}$ are symmetric, homogeneous,
and short-ranged in the index space, one can approximate the last term by
a second-order differential and derive Eq.~(\ref{eq:rd})
through a suitable limiting process~\cite{ebeling}.

It has been suggested in
Ref.~\onlinecite{inno} that $U(x,t)$ can be identified with the cumulative
distribution $C(x,t) \equiv \int_{-\infty}^x P(x',t) dx'$ up to a
proportionality coefficient absorbed into $k$, because
those with $x'<x$ are potential adopters of $x$.
In the general context of fads, however, the payoff function might not be such a
simple linear function of $C(x,t)$. For example, there can be an adverse
effect of being too popular.
Let us generalize the above argument in the following way:
We assume that there eventually develops a traveling wave
with speed $v$ and a unimodal shape of $P(x-vt)$.
At any time, the corresponding $C(x)$ should be a
monotonically increasing smooth function of $x$ within a certain region. Our
claim is that it can then be used as a basis for expressing an arbitrary shape
of $U(x)$ within the region of our interest. The idea is as follows:
If we define $\mu \equiv C(x)$, there is one-to-one correspondence between $x$
and $\mu$ as long as $P>0$, and one can write $U$ as a function of
$\mu$ instead of $x$. We expand $U(\mu)$ as a polynomial in $\mu$, i.e.,
$U = a_0 + a_1 \mu + a_2 \mu^2 + \cdots = a_0 + a_1 C(x) + a_2 C^2(x) +
\cdots$ with coefficients $a_i$. Clearly, $a_0$ does not alter the
dynamics as $\left< U \right>$ is subtracted, and $a_1$ is absorbed
into the equation by rescaling the time scale.
We will therefore retain the two lowest-order
contributions as $U(x,t) = C(x,t) + \frac{3}{2}g C^2(x,t)$ with a
constant $g$, where the factor of $\frac{3}{2}$ is added for later convenience.
This constant $g$ is introduced as a shape parameter to
control the skewness of the distribution, but it
can also be interpreted as the population's
preference to new fads. If $g$ is negative, for example,
it means that the population is `conservative' in the sense that the fitness
$U$ does not increase with $x$ as sharply as it would under positive $g$.
Noting that
\begin{equation}
\int_{-\infty}^{\infty} C^n Pdx = \int_0^1 C^n dC = \frac{1}{n+1},
\end{equation}
we rewrite Eq.~(\ref{eq:rd}) as
\begin{equation}
\frac{\partial P}{\partial t} = k \left[ C+ \frac{3}{2}g C^2 -
\frac{1}{2}(1+g) \right] P + D \frac{\partial^2 P}{\partial x^2}.
\label{eq:original}
\end{equation}
By rescaling $t$ and $x$ as $k t$ and $x\sqrt{k/D}$, respectively, we reduce
the equation to
\begin{equation}
\frac{\partial P}{\partial t} = \left[ C+ \frac{3}{2}g C^2 - \frac{1}{2}
(1+g) \right] P + \frac{\partial^2 P}{\partial x^2}.
\end{equation}
By using $\frac{\partial C}{\partial x}
= P$, this can be written as
\begin{equation}
\frac{\partial^2 C}{\partial t \partial x}
= \frac{\partial}{\partial x} \left[ \frac{C^2}{2} + \frac{g}{2} C^3 -
\frac{1}{2} (1+g) C + \frac{\partial^2 C}{\partial x^2} \right],
\end{equation}
which implies that
\begin{equation}
\frac{\partial C}{\partial t} = \frac{1}{2} C(C-1) ( g C + g+1) 
+ \frac{\partial^2 C}{\partial x^2} + \sigma(t),
\end{equation}
where $\sigma(t)$ is a function of $t$ only. Substituting $x=\pm \infty$,
we find that $\sigma(t) = 0$ because the left-hand side and the first two
terms in the RHS identically vanish there.
To sum up, we will consider the following equation:
\begin{equation}
\frac{\partial C}{\partial t} = \frac{1}{2} C(C-1) (g C + g+1)
+ \frac{\partial^2 C}{\partial x^2}.
\label{eq:gov}
\end{equation}
This form is close to the Zeldovich equation~\cite{zeldo},
although the phase-space structure is different as will be detailed below.
If $g=0$, this is mathematically equivalent to the celebrated Fisher
equation~\cite{fisher}. This model can also be interpreted in terms of the
activator-inhibitor model in Ref.~\onlinecite{given} (see
Appendix~\ref{sec:aim} for details).
In addition,
one may recall that a spin system is phenomenologically expressed by the
time-dependent Ginzburg-Landau equation as
\begin{equation}
\frac{\partial m}{\partial t} = m-m^3  + \nabla^2 m,
\end{equation}
where $m$ denotes the order parameter~\cite{kin}. One keeps the linear
and cubic terms of $m$ to describe the symmetric double-well structure
of the free-energy density functional. In this respect, the additional term
proportional to $g$ in deriving Eq.~(\ref{eq:gov}) can be interpreted
as introducing a cubic term to the RHS,
although we are not dealing with a magnetic system. 

The assumption of the traveling wave implies that
$C(x,t) = C \left(x-vt \right)$, where $v$
is the speed. In a moving frame with coordinate $\eta \equiv x-vt$,
Eq.~(\ref{eq:gov}) describes the shape of the wave as a second-order ordinary
differential equation:
\begin{equation}
-v \frac{dC}{d\eta} = \frac{1}{2} C(C-1) ( g C +g+1 )
+ \frac{d^2 C}{d \eta^2},
\label{eq:ode}
\end{equation}
or, equivalently, a set of two first-order ordinary differential equations:
\begin{equation}
\left\{
\begin{array}{lcl}
dC/d\eta &=& P\\
dP/d\eta &=& -vP - \frac{1}{2} C(C-1) (g C +g+1),
\end{array}
\right.
\label{eq:set}
\end{equation}
where the first line relates the probability density function $P$
and its cumulative $C$.
This system has three fixed points at $(C,P)=(0,0)$, $(1,0)$, and
$(-1-g^{-1},0)$.
The linear
stability analysis shows that the eigensystem around
the origin consists of eigenvalues $\lambda_\pm = \frac{1}{2} \left[-v \pm
\sqrt{v^2 + 2+2g} \right]$ and eigenvectors
\begin{equation}
\mathbf{y}_\pm = \left(
\begin{array}{c}
1 \\ \lambda_\pm
\end{array}
\right).
\end{equation}
On the other hand, the eigensystem in the vicinity of $(C,P)=(1,0)$
is obtained as $\mu_\pm = \frac{1}{2} \left(-v \pm \sqrt{v^2-4g -2}
\right)$, together with the corresponding eigenvectors:
\begin{equation}
\mathbf{z}_\pm = \left(
\begin{array}{c}
1 \\ \mu_\pm
\end{array}
\right).
\end{equation}
If $v < \sqrt{4g+2}$, the eigenvalues become complex, which
is not physically reasonable because $C$ would not be bounded within the
unit interval. This proves the existence of the minimum speed
$v_{\rm min} = \sqrt{4g+2}$ below which one cannot find a
traveling-wave solution
(see Appendix~\ref{sec:speed} for more discussion and references on the
speed selection principle).
If $g = -\frac{1}{2}$, the wave can stop, meaning that the population is
so conservative that there might be no progression to higher $x$.
The last fixed point at $(C,P)=(-1-g^{-1},0)$ has an eigensystem
of $\zeta_\pm = \frac{1}{2} \left[ -v \pm \sqrt{v^2 - 4g^2-6-2g^{-1} 
} \right]$ and
\begin{equation}
\mathbf{w}_\pm = \left(
\begin{array}{c}
1 \\ \zeta_\pm
\end{array}
\right).
\end{equation}
The last fixed point should not lie between the other two,
because it is meant to modify the flow from $(0,0)$ to $(1,0)$,
leaving the overall phase-space structure unaltered.
This restricts the possible range of $g$ to
$g \ge -\frac{1}{2}$.
\begin{figure}
\includegraphics[width=0.45\textwidth]{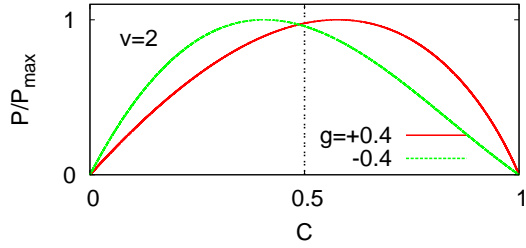}
\caption{(Color online) $P$ versus $C$, where $P$ is scaled by
its maximum amplitude $P_{\rm max}$. The speed is chosen as $v=2$.
Note that the flow of Eq.~(\ref{eq:set})
goes from $(C,P)=(0,0)$ to $(1,0)$ as $\eta \equiv x-vt$ increases. (a)
The solid line with $g=+0.4$ has $\frac{dP}{dC}>0$ at $C=\frac{1}{2}$ in this
plot, and $P(\eta)$ has negative skewness. If we fix $x$
and observe its time series as $t$ varies, therefore,
the height grows rapidly and then decreases slowly.
(b) The dotted line with $g=-0.4$ shows the opposite case, as explained
in the main text.
}
\label{fig:skew}
\end{figure}
With this parameter $g$, one can control the skewness of
the wave shape:
Consider a trajectory from $(0,0)$ to $(1,0)$ in the $(C,P)$ plane,
as described by Eq.~(\ref{eq:set}).
If the shape of $P(\eta)$ is negatively skewed,
we will see $\frac{dP}{dC} > 0$ at $C=\frac{1}{2}$, and vice versa
(Fig.~\ref{fig:skew}).
Setting $C=\frac{1}{2}$, we see from Eq.~(\ref{eq:set}) that
\begin{equation}
\frac{dP}{dC} = -v + \frac{1 + 3g/2}{8P^\ast},
\label{eq:dpdc}
\end{equation}
where $P^\ast$ is the value of $P$ when $C=\frac{1}{2}$.
Equation~(\ref{eq:dpdc}) shows that one can change the sign of
$\frac{dP}{dC}$, hence the skewness, by modulating $g$.
It is not a linear relationship, though, because
$P^\ast$ also depends on $v$ and $g$.
A negative value of $g$ close to its lower bound $-\frac{1}{2}$
will lead to a positively skewed shape of $P(\eta)$ (see Fig.~\ref{fig:skew}).
As this wave travels in time, we will see that
fads are adopted slowly and then abandoned rapidly, compared with the
case of positive $g$.

\section{Data Analysis}
\label{sec:data}

\begin{figure}
\includegraphics[width=0.45\textwidth]{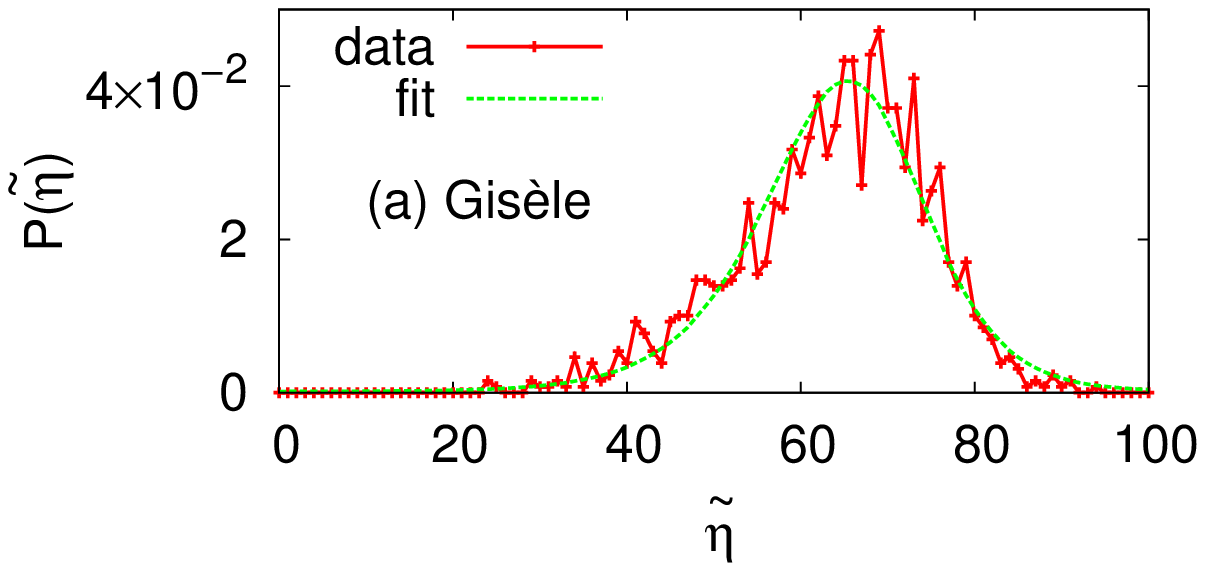}
\includegraphics[width=0.45\textwidth]{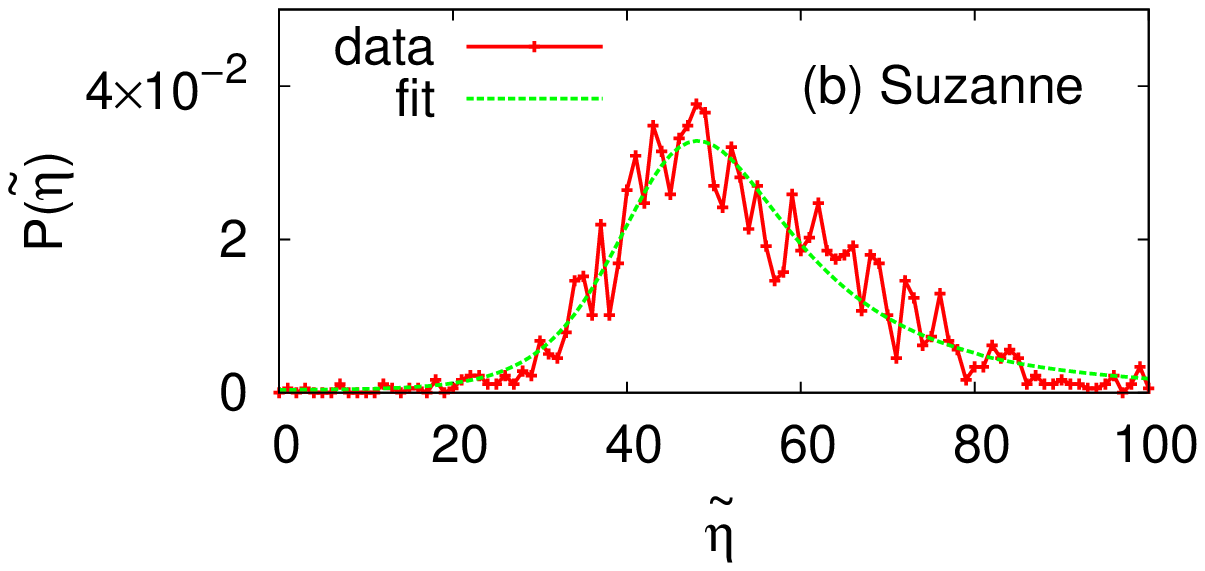}
\includegraphics[width=0.45\textwidth]{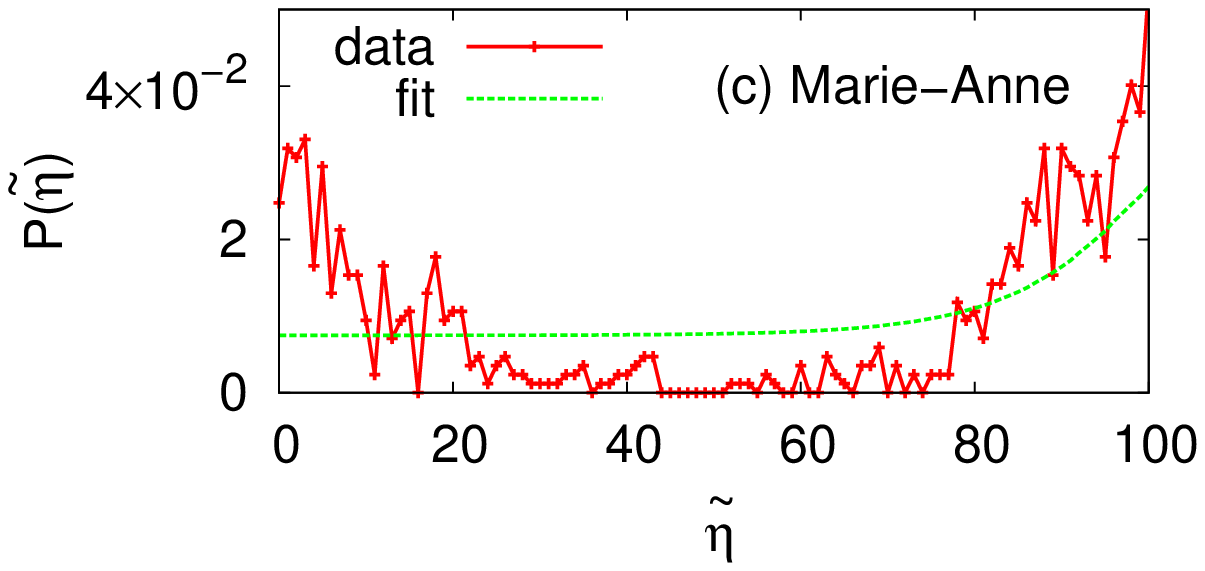}
\includegraphics[width=0.45\textwidth]{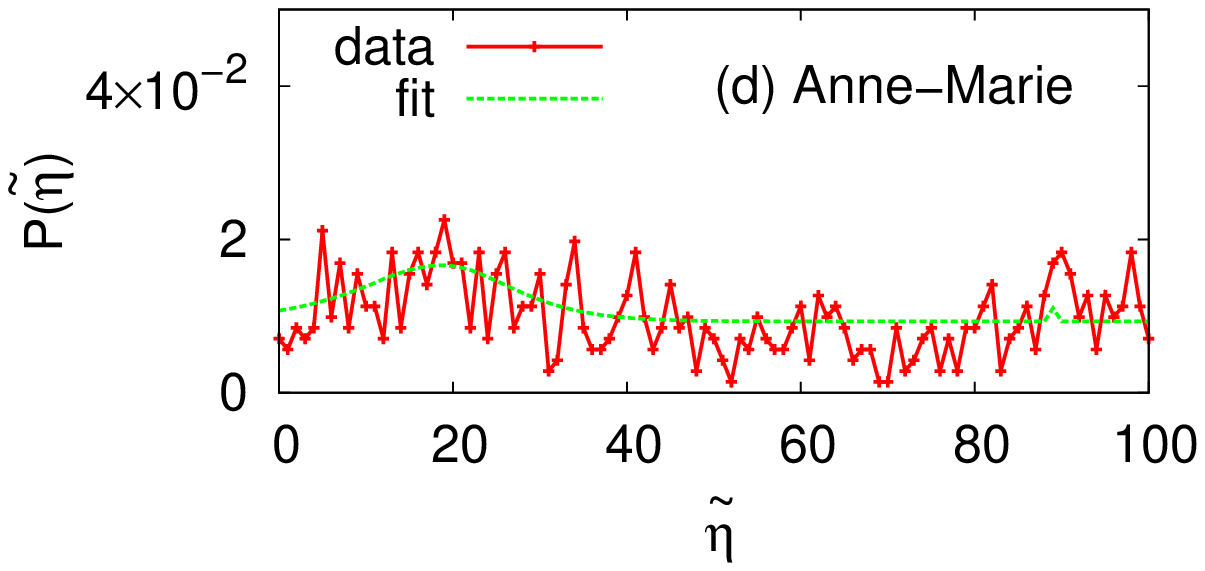}
\caption{(Color online)
(a) More than two-thirds of the names show left-skewed $P(\tilde\eta)$, and
one such example is plotted here.
It means that they quickly become popular and then slowly disappear,
because $\tilde\eta \propto -t$.
The noisy curve represents the corpus data,
and the smooth one is obtained from our model.
This example is fitted with $v=1.7$ and $g=-0.06$ (see the main text for
details of the fitting).
(b) A dozen of names are equally well or better fitted to
right-skewed distribution. The fitting
parameters are $v=2.4$ and $g=-0.43$ in this example.
(c) Recurrence is
observed for a couple of names like Gabrielle and Marie-Anne,
roughly with a centennial period.
(d) There are a few `steady sellers', including Rachel and {\'E}lisabeth,
with no particular patterns.}
\label{fig:qfem}
\end{figure}

For given $v$ and $g$, we can obtain $P(\eta) = P(x-vt)$
by numerically integrating Eq.~(\ref{eq:set}). It is \emph{not}
directly observable, however,
because it is difficult to measure $x$ in practice.
The available data set is given as a collection of time series, one for each
different given name.
If we choose one of them, $x$ is fixed by assumption,
and the observed time series corresponds to $P(t-x/v)$. That is,
if compared with the hypothetical distribution $P(\eta)$,
the time series would be a reflected and scaled image along the horizontal axis.
Let us stress the underlying assumptions behind this statement:
First, the parameters such as $v$ and $g$ are assumed to be stationary,
at least approximately, to relate the observed time series to $P(\eta)$.
Second, our model assumes that every name is equal, except for its value $x$.
Therefore, in principle, every name is expected to exhibit a similar pattern
in its time series. Recall that each time series should be related to others
by time translation, because it appears as $P(t-x/v)$.
For each time series to be regarded as a projected image of
a probability distribution function, therefore,
our data should be traced backward
with normalizing the area under the curve to one.
In other words, the time series $f(t)$ up to $t=t_{\rm max}$ transforms to
$f(t_{\rm max}-t) / \sum_t f(t)$, which we denote as $P(\tilde\eta)$
with $\tilde\eta \equiv t_{\rm max}-t$.

Every year from 1900 to 2000, we randomly sample $2 \times 10^3$ people out
of the corpus and will work with this sampled data henceforth.
Otherwise, it would be difficult to compare the number of people given
a particular name, counted over the century, with that of another name,
because the size of the corpus varies year by year.
The resulting $P(\tilde\eta)$ for the Quebec female names
shows four representative types:
The first is negatively skewed as depicted in Fig.~\ref{fig:qfem}(a),
and this comprises the major part of the hundred names.
We have found negative skewness for about $60$ names, and the actual number
could well be greater than this, because very old or new names show only one
of the tails in their time series so that their skewness cannot be determined
precisely.
There are less than ten names that are better described by
right-skewed distribution [Fig.~\ref{fig:qfem}(b)]. Only a couple
of names exhibit such recurrent behavior [Fig.~\ref{fig:qfem}(c)]
as has been considered in Ref.~\onlinecite{kessler}.
Note that the recurrence is entirely ignored in our model.
Finally, we find three names that are noisy around their respective mean
values, and thus not covered by any of the categories above
[Fig.~\ref{fig:qfem}(d)].

The sampled data points are fitted to our model curve, integrated from
Eq.~(\ref{eq:set}). The objective is to minimize the following cost function:
\begin{equation}
\chi^2 = \sum_i \frac{(n_i - Np_i)^2}{Np_i},
\label{eq:chi2}
\end{equation}
where $n_i$ is the number of people given a particular name at year $i$,
$N$ is the total number of people given this name over the period from 1900
to 2000, and $p_i$ is the predicted fraction from our model.
We interpret this problem as distributing $N$ people into 101 bins from 1900 to
2000.
We should have four fitting parameters, two of which are $v$ and $g$,
and the other two determine translation and scaling along the horizontal axis.
We have added one more for a vertical shift,
because the $\chi^2$ statistic [Eq.~(\ref{eq:chi2})]
is too sensitive to background noise:
Our model describes how a fad of a name catches on and fades away over
a certain period,
so it has $p_i \approx 0$ outside that period. In practice, however,
the name can still be found as a stochastic effect. If $p_i$ is vanishingly
small there, the cost function $\chi^2$ diverges. If the noisy fluctuations are
not taken into account, therefore,
the fitting program will try to avoid this divergence
at every cost, even if it misses the main signal.
The number of degrees
of freedom thus amounts to $d_f = 101-1-5=95$. The rule of thumb
is that $\chi^2$ should be of $O(d_f)$. The conventional criteria
for statistical significance allow $\chi^2$ to be roughly as high as $130$,
which is observed with probability $0.01$ in the $\chi^2$ distribution.

\begin{figure}
\includegraphics[width=0.45\textwidth]{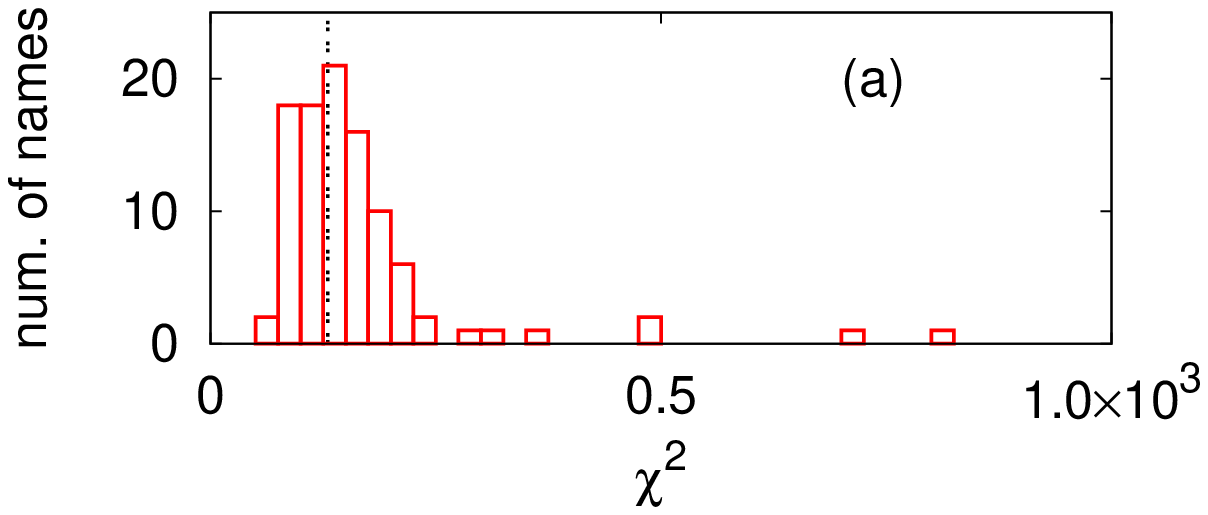}
\includegraphics[width=0.45\textwidth]{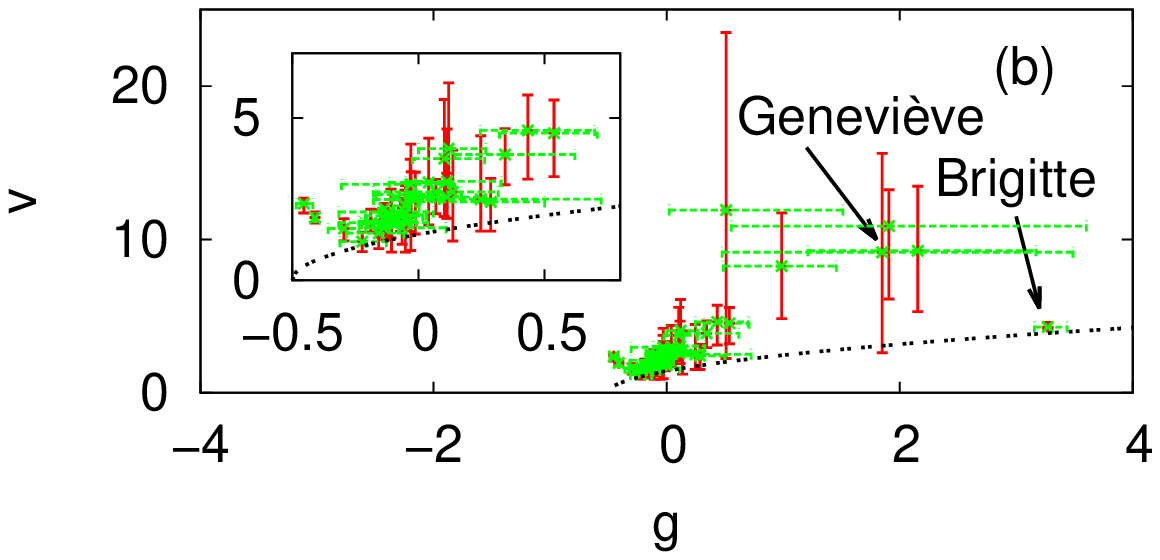}
\includegraphics[width=0.45\textwidth]{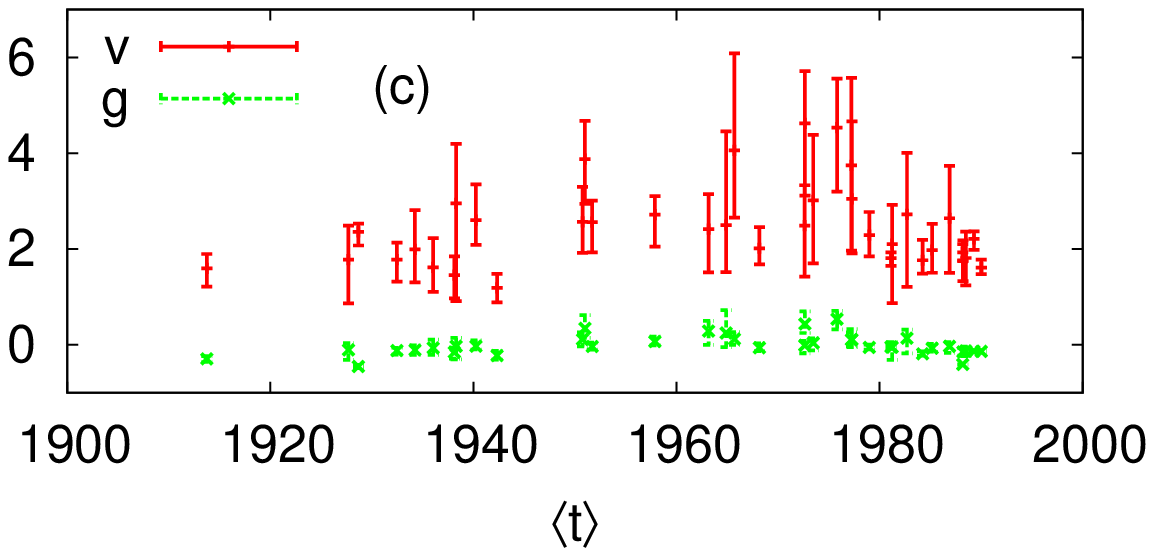}
\includegraphics[width=0.45\textwidth]{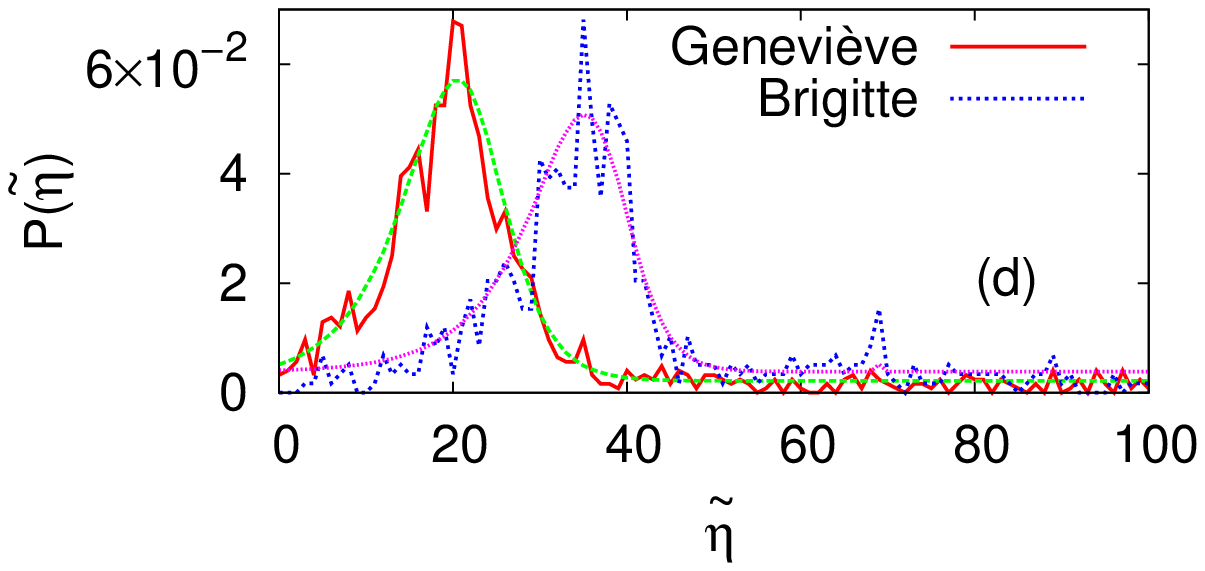}
\caption{(Color online)
Monte Carlo fitting results with the Metropolis algorithm.
The energy function is identified with $\chi^2$ and the temperature is set to
be $T=1$. (a) The distribution of
$\chi^2$ [Eq.~(\ref{eq:chi2})] for the hundred female names after the fitting.
The vertical line means $\chi^2=130$, below which the statistical significance
is greater than $0.01$. (b) Distribution of
$v$ and $g$ for the names fitted with $\chi^2 < 130$. The error bars express
the ranges of the parameters during $10^5$ Monte Carlo steps. Inset: A zoomed
view. The dotted lines represent $v_{\rm min} = \sqrt{4g + 2}$.
(c) Estimated $v$ and $g$
against the mean time of prevalence [Eq.~(\ref{eq:meant})] for the names
in the inset of (b).
(d) Actual data and fitting curves for two outliers on the $(g,v)$ plane,
indicated by the arrows in (b).
}
\label{fig:chi2}
\end{figure}

The parameters are adjusted by the Metropolis algorithm, in which
Eq.~(\ref{eq:chi2}) plays the role of the energy function.
The energy landscape will
be simple enough to apply the zero-temperature Monte Carlo method, as
long as the initial condition is reasonably chosen.
To estimate the possible ranges of the fitting parameters, however, we
choose a finite temperature $T=1$ because fluctuations of $O(1)$ in
$\chi^2$ will hardly affect the statistical significance.
The initial parameters for translation and scaling can be estimated
by checking the statistical moments of the data, and we try two initial
values of $g=0$ and $g=0.4$, with $v = v_{\rm min}(g)$, for faster
convergence. The resulting distribution
of $\chi^2$ for the hundred female names is plotted in Fig.~\ref{fig:chi2}(a).
The figure shows that about 40 percents of the hundred female names are
successfully described by our model, yielding $\chi^2 \lesssim 130$.
For these names, we also plot the ranges of their fitting parameters
in Fig.~\ref{fig:chi2}(b).
Here, one immediately finds a cluster of names with $g \approx 0$ and
$v \lesssim 5$, which is separately depicted in the inset.
We see from this figure that many of the parameter
values do overlap as predicted by our theory.
It is also interesting
that the estimated speed is not far from $v_{\rm min}$, if we consider that
a new fad would start with a sharp initial condition.
Figure~\ref{fig:chi2}(c) shows the estimated parameters of the names
inside the cluster, against their $\left<t \right>$ [Eq.~(\ref{eq:meant})].
It confirms that the parameters have been stationary throughout
the period of observation, which is one of our underlying assumptions.

We also have a few outliers that
accept fairly large $v$ or $g$. Let us take a closer look
by plotting the data for two of them, Genevi\`eve and
Brigitte [Fig.~\ref{fig:chi2}(d)]. Then, we immediately find that the
large $v$ of the former name is likely to be an artifact
due to the steady noise level, because a broader distribution
may be interpreted as faster propagation (see Appendix~\ref{sec:speed}).
The same explanation applies to
all the other names with large error bars in Fig.~\ref{fig:chi2}(b).
On the other hand,
the latter case of Brigitte shows exceptionally large skewness in spite
of rather stable parameter estimations [Fig.~\ref{fig:chi2}(b)].
Such a distorted shape is probably
caused by external factors that our model cannot account for.

Overall,
we conclude that our model does explain a part of the observed data,
because it describes about 40 names with consistent values of $v$ and $g$.
At the same time, the degree of heterogeneity among the given names
has turned out to be higher than we expected from Fig.~\ref{fig:wave}.

\section{Discussion and Summary}
\label{sec:discussion}

In summary, we have constructed a model to describe a pattern of fads.
Our working
hypothesis has been that although each fad is unpredictable, the overall
pattern tends to repeat itself over and over again.
Figure~\ref{fig:wave} provides a striking indication of this hypothesis,
and suggests that the governing dynamics can develop a traveling-wave solution
with a well-defined shape. We have built up a mean-field theory of selection
and mutation to explain the observation in Fig.~\ref{fig:wave}.
The assumption of a traveling-wave solution implies that the time series of
every name can be fitted to a single curve if translated horizontally.
We have introduced a shape parameter $g$ in addition to the wave speed $v$,
and checked whether these parameters give consistent estimates over the
hundred time series.
Our model must have oversimplified the reality as we all know that
the naming dynamics is involved with many accidental factors such as
celebrities and mass media. So as long as the theory predicts sensible behavior,
the question should be how much these random fluctuations alter the dynamics.
Our analysis suggests the existence
of the regularity on a qualitative level:
Most of the names are described by unimodal left-skewed
distribution, which means that it is common to adopt a fad rapidly and then
abandon it slowly. Quantitatively,
we have found meaningful fitting results for about $40$ names among $100$.
Since we have missed more than a half, we also conclude that the
heterogeneity is not so negligible as assumed in the theory.

\acknowledgments
We are grateful to L. Duchesne for providing us with the data.
S.K.B. was supported by Basic Science Research Program through the
National Research Foundation of Korea funded by the Ministry of Science,
ICT and Future Planning (NRF-2014R1A1A1003304).
B.J.K. was supported by the National Research Foundation of Korea grant
funded by the Korea government (MSIP) (No. NRF-2014R1A2A2A01004919).
This work was supported by the Supercomputing Center/Korea Institute of
Science and Technology Information under Project No. KSC-2014-C1-004.

\appendix

\section{Activator-inhibitor model}
\label{sec:aim}

Reference~\onlinecite{given} suggests an alternative model to explain the
empirical
pattern of given names as follows: Consider a particular name and let $P(t)$ and
$Q(t)$ be the fraction of couples that give the name to their babies and the
fraction of people that bear the name at time $t$, respectively. The former
fraction of couples `activate' the use of this name, whereas the latter fraction
of people tend to `inhibit' it.
If $\kappa(t)$ means the mortality rate at $t$, the change of
$Q(t)$ over a time scale $\tau$ can be written as
\begin{equation}
\frac{d}{dt}Q(t) = \tau^{-1} P(t) - \kappa(t) Q(t),
\end{equation}
with a formal solution:
\begin{equation}
Q(t) = \tau^{-1} e^{-I(t)} \int_{-\infty}^t P(t') e^{I(t')} dt' + Q(-\infty)
e^{-I(t)},
\label{eq:b}
\end{equation}
where $I(t) \equiv \int_{-\infty}^t \kappa(t') dt'$. The second term
on the RHS can be discarded by setting the surface term at
$t=-\infty$ as zero.
On the other hand, the governing equation for $P(t)$ is assumed to be
\begin{equation}
\frac{d}{dt}P(t) = -\alpha P^l(t) + \beta \left[ 1 - \frac{P(t)}{P_s} \right]
\left[ 1 - \frac{Q(t)}{Q_s} \right] P(t),
\end{equation}
where $\alpha$, $\beta$, $l$, $P_s$, and $Q_s$ are model parameters. The first
term on the RHS describes a threshold effect, and $P_s$ and $Q_s$
mean saturation levels for $P$ and $Q$, respectively.

In this appendix, we compare this activator-inhibitor model with our approach.
Let us first assume that the threshold phenomena are negligible and
that $P \ll P_s$ all the time. Then, the equation simplifies to
\begin{equation}
\frac{d}{dt}P(t) = \beta \left[ 1 - \frac{Q(t)}{Q_s} \right] P(t).
\end{equation}
If the mortality rate is small enough within the time scale of observation,
we can approximate Eq.~(\ref{eq:b}) as
\begin{equation}
Q(t) \approx \tau^{-1} \int_{-\infty}^t P(t') dt'.
\label{eq:Qapprox}
\end{equation}
Our model considers a traveling wave of velocity $v>0$ represented by $P(x,t) =
P(x-vt)$. It implies that the following integral over time 
\begin{equation}
\int_{-\infty}^t P(x-vt') dt' = -v^{-1} \int_{-\infty}^{x-vt} P(x-vt') d(x-vt')
\end{equation}
is related to the cumulative distribution
\begin{equation}
C(x-vt) = \int_{-\infty}^x P(x'-vt) dx' = \int_{-\infty}^{x-vt} P(x'-vt)
d(x'-vt).
\end{equation}
Roughly speaking, therefore, it is $1-C(x,t)$ that plays the role of inhibition
from those who bear the name in our model.
Plugging this into Eq.~(\ref{eq:Qapprox}), we see that the increase
of $P$ is proportional to $C$, which can be written as
\begin{equation}
\frac{\partial P}{\partial t} = k \left( C - \left< C \right> \right) P,
\label{eq:logistic}
\end{equation}
where $k$ is a proportionality constant,
and we need $\left< C \right> = \int_{-\infty}^{\infty}
C(x',t) P(x',t) dx'$ in the parentheses to guarantee the conservation of total
probability $\int_{-\infty}^{\infty} P(x',t) dx' = 1$. The constant $k$ can
be absorbed into $t$ by rescaling the time scale.
Reference~\onlinecite{inno} shows
that Eq.~(\ref{eq:logistic}) has a solution
\begin{equation}
C(x,t) = \frac{1}{2} \tanh \left[h(x) - \frac{t}{4} \right] + \frac{1}{2}
\end{equation}
with a certain function $h(x)$ such that $dh/dx \ge 0$, $h(x \rightarrow
+\infty) = +\infty$, and $h(x \rightarrow -\infty) = -\infty)$.
If this is a traveling wave with a constant velocity, $h(x)$ should be a
linear function and $P(x,t)$ thus has a peak with zero skewness.
Reference~\onlinecite{inno} has found that we can make it skewed just
by adding a diffusive term to the `reaction' between activators and
inhibitors that Eq.~(\ref{eq:logistic}) describes.
The present work has generalized this finding by introducing a shape parameter
called $g$ to control the skewness. It is straightforward to see that
Eq.~(\ref{eq:logistic}) is obtained from Eq.~(\ref{eq:original})
by taking $g=0$, $k=1$, and $D=0$.
The activator-inhibitor model in Ref.~\onlinecite{given} differs
from our reaction-diffusion approach in that it mainly focuses on
elaborating the reaction kinetics, e.g., by including threshold and saturation
effects for $P$. In contrast to such single-name dynamics,
the diffusive process introduces other names to the population,
playing a similar role to mutation in biology.

\section{Speed selection principle}
\label{sec:speed}
How fast the traveling wave propagates into a linearly unstable state is
one of main concerns when we study a reaction-diffusion system.
This is usually known as the front propagation problem.
In the context of the original Fisher equation~\cite{fisher},
the traveling wave describes expansion of a population, whose
density is denoted as $u(r, t)$,
where $r$ means a one-dimensional real space.
Under an assumption that $u$ spreads with keeping a constant shape
at an asymptotic speed $v_{\rm as}$ in the large-$t$ limit,
$u(r, t)$ can be approximated as $u(r-v_{\rm as}t)$.
The speed $v_{as}$ of the nonlinear wave is governed by its
initial profile at $t=0$ and the leading
edge~\cite{kpp, larson, mckean, rothe, saarloos}.
Let us consider a monotonically decreasing wave profile, such as $u \sim
e^{-\zeta r}$. It is known that
there is a critical value $\zeta = \zeta^\ast$, above which
the wave propagates with the lowest speed $v^\ast$. Therefore,
if the initial profile is steeper than $e^{-\zeta^\ast r}$,
the wave moves forward at the asymptotic speed $v_{as}$
equal to $v^\ast$.
Otherwise, it travels at $v_{as}$ larger than $v^\ast$,
and $v_{as}$ is controlled by the dynamics of the edge which is
obtained by linearizing the original equation.

In our notation,
$1-C$ maps to the population density $u$ in the Fisher equation.
As far as a well-mixed population is concerned,
the spatial dependency is negligible and we have $u(r, x, t)=u(x, t)$.
For finding the minimum propagation speed $v^\ast$,
let us substitute $u=1-C$ into Eq.~(\ref{eq:gov}) to obtain
\begin{equation}\label{eq:linear1}
\frac{\partial u}{\partial t}=
\frac{1}{2}u(1-u)(1+2g-2u)+\frac{\partial^2 u}{\partial x^2}.
\end{equation}
Equation~(\ref{eq:linear1}) is linearized at the edge ($u\ll1$),
which results in
\begin{equation}\label{eq:linear2}
\frac{\partial u}{\partial t}
=\frac{1}{2}u(1+2g)+\frac{\partial^2 u}{\partial x^2}.
\end{equation}
Suppose that the wave profile has an exponential tail in the $x$ space,
i.e., $u(x, t)=u(x-vt)\sim e^{-\zeta(x-vt)}$.
If we plug this ansatz into Eq.~(\ref{eq:linear2}),
it yields a quadratic equation for $\zeta$:
\begin{equation}
\zeta^2-v\zeta+\frac{1}{2}(1+2g)=0.
\end{equation}
Solving this equation, one obtains a dispersion relation for
the edge speed as a function of $\zeta$ as follows:
\begin{equation}\label{eq:linear3}
v(\zeta)=\zeta+\frac{1}{2\zeta}(1+2g).
\end{equation}
This expression diverges as $\zeta\rightarrow0$
or $\zeta\rightarrow\infty$ and always has a minimum
at $\zeta^\ast$ between the two regimes.
The derivative of Eq.~(\ref{eq:linear3}) with respect to $\zeta$
vanishes at such $\zeta^\ast$:
\begin{equation}\label{eq:linear4}
\frac{\partial v(\zeta)}{\partial \zeta}=1-\frac{1+2g}{\zeta^{2}} = 0,
\end{equation}
which has two extrema at $\zeta=\pm \sqrt{(1+2g)/2}$.
The one with the positive sign gives the minimum speed:
\begin{equation}\label{eq:linear5}
v^\ast=\sqrt{2(1+2g)}.
\end{equation}
Then, $\zeta^\ast$ can be expressed self-consistently in terms of $v^\ast$,
i.e., $\zeta^\ast=v^\ast/2$.

%
\end{document}